\documentclass[aps,prb,reprint,superscriptaddress]{revtex4-2}
\usepackage{graphicx} 
\usepackage[T1]{fontenc}
\usepackage{amsmath} 
\usepackage{bm} 
\usepackage{natbib}
\usepackage{natmove}
\usepackage[colorlinks=true, linkcolor=blue, citecolor=blue, urlcolor=blue, linktoc=page, bookmarks=false, pdfstartview={FitH}, pdfborder={0 0 0.0 [3 3]}]{hyperref} 
\usepackage{color} 
\usepackage{dcolumn} 
\newcolumntype{.}{D{.}{.}{1.3}}
\newcolumntype{-}{D{.}{.}{4.0}}
\newcolumntype{:}{D{:}{:}{12.30}}
\newcolumntype{*}{D{:}{:}{1.7}}
\usepackage{makecell}
\usepackage{multirow}
\usepackage{cleveref}
\usepackage{easyReview} 
\crefname{figure}{Fig.}{Figs}
\crefname{table}{Table}{Tables}
\crefname{section}{Sec.}{Secs.}
\crefname{equation}{Eq.}{Eqs.}
\newcommand{\blue}{\color{blue}}

\renewcommand{\today}{\number\day \space \ifcase \month \or January\or February\or March\or April\or May\or June\or July\or August\or September\or October\or November\or December\fi \space \number\year} 
\def\m1r{\multicolumn{1}{r}}
\begin{document}
\title{Quantum Geometric Origin of N\'eel-Order-Driven Magneto-Optical Response in MnTe}
\author{Suman \surname{Rooj}}
\email[Email: ]{suman.rooj@physics.uu.se}
\affiliation{Department of Physics and Astronomy, Uppsala University, P. O. Box 516, SE-75120 Uppsala, Sweden}

\author{M. S. \surname{Mrudul}}
\affiliation{Department of Physics and Astronomy, Uppsala University, P. O. Box 516, SE-75120 Uppsala, Sweden}
\author{Daegeun \surname{Jo}}
\affiliation{Department of Physics and Astronomy, Uppsala University, P. O. Box 516, SE-75120 Uppsala, Sweden}
\affiliation{Wallenberg Initiative Materials Science for Sustainability, Uppsala University, SE-75120 Uppsala, Sweden }
\author{Peter M. \surname{Oppeneer}}
\email[Email: ]{peter.oppeneer@physics.uu.se}
\affiliation{Department of Physics and Astronomy, Uppsala University, P. O. Box 516, SE-75120 Uppsala, Sweden}
\affiliation{Wallenberg Initiative Materials Science for Sustainability, Uppsala University, SE-75120 Uppsala, Sweden }
\date{\today}
\begin{abstract}
Magneto-optical (MO) responses provide a direct probe of magnetic order, yet resolving compensated antiferromagnetic order remains challenging. Here, we demonstrate in altermagnetic MnTe that the symmetry of the N\'{e}el order is imprinted on the MO response through distinct quantum-geometric contributions. Combining first-principles calculations with a Landau symmetry framework, we demonstrate that the quantum metric gives rise to a twofold optical anisotropy, with its diagonal and symmetric off-diagonal components forming an $E_{2g}$ doublet associated with the quadratic N\'{e}el-order invariants, providing a characteristic MO Voigt signature of the N\'{e}el-vector orientation. In contrast, the Berry curvature generates a complementary threefold Hall response governed by the cubic $A_{2g}$ N\'{e}el-order invariant. Although the Hall conductivity and the out-of-plane weak-ferromagnetic moment share the same symmetry and angular dependence, constraining the latter to zero leaves the optical Hall response essentially unchanged demonstrating that the Hall response originates directly from the compensated N\'{e}el order rather than the net magnetization. Our results establish quantum geometry as a direct link between N\'{e}el order and MO response, revealing distinct twofold and threefold optical signatures of N\'{e}el order in altermagnetic MnTe.
\end{abstract}
\maketitle
{\it{Introduction}}--- Magneto-optical (MO) phenomena provide a powerful means of probing magnetic order through the interaction between light and the electronic degrees of freedom \cite{Zvezdin1997,OppeneerBook}. In conventional ferromagnetic materials, light's polarization rotation and ellipticity are commonly associated with a net magnetization and the resulting time-reversal-symmetry-breaking optical Hall response \cite{OppeneerBook}. This picture, however, becomes less straightforward for compensated antiferromagnets (AFMs), where the magnetic order is characterized by a N\'{e}el vector rather than a net magnetization. Nevertheless, the underlying magnetic order can leave distinct signatures in the MO response, providing a route to resolve compensated magnetic order without relying on a net moment \cite{Ferre1984,Oppeneer2003,Saidl2017optical,Nemec2018antiferromagnetic,Kimel2019,Pan2026NeelMOKE,Sunko2026MOKE}.  Understanding how the different components of the MO response encode the underlying magnetic order is therefore important both for interpreting existing MO measurements and for identifying new optical probes of compensated magnetism.

Altermagnetic materials offer a compelling platform in which magnetic order can gives rise to a range of unconventional electronic responses despite vanishing net magnetization \cite{LiborPRX2022,Song2025Altermagnets,LiborSciAdv2020, Libor2022,Zunger2022, YuanPRM21}. In an altermagnet (AM), the compensated magnetic structure is accompanied by momentum-dependent spin splitting dictated by crystal and magnetic symmetries \cite{Zhang2007,LiborPRX2022,Libor2022,YuanPRM21}. Hexagonal MnTe is a prototypical realization~\cite{Krempasky2024,MnTeArpes2024,MnTePRB2024,Mazin2023, RoojMnTe2024}, exhibiting unconventional MO and other transport phenomena~\cite{MnTeAHE2023,MnTeAHE2024,MnTeXMCD2024,GonzalezBetancourt2024,Ultrafast2026,MO2026}. In particular, a recent study~\cite{Ultrafast2026} has demonstrated that the reflection MO Voigt effect (MOVE) \cite{Voigt1908magneto,Note,Oppeneer2001,Valencia2010} and the MO Kerr effect (MOKE) in MnTe serve as distinctive fingerprints of altermagnetism: the symmetric conductivity governs the anisotropic optical response associated with the MOVE, while the antisymmetric component gives rise to the Hall-like optical response underlying the MOKE. These observations propose the potential of MO phenomena to probe compensated magnetic order through both symmetric and antisymmetric components of the optical response. Their distinct dependence on the N\'{e}el order therefore provides a route to optically resolve its symmetry. However, a fundamental understanding of how these distinct optical channels are governed by the symmetry of the N\'{e}el order remains elusive. 

Quantum geometry provides a natural framework for connecting magnetic symmetry with optical response. It characterizes the geometry of Bloch states through the quantum metric and Berry curvature, which capture complementary aspects of their momentum-space structure and govern various electronic and optical phenomena \cite{Ahn2022,YuQGT2025,LiuQGT2025,JiangQGT2025}. Although quantum geometry has attracted considerable attention in the context of nonlinear transport \cite{sala2025}, it also underlies linear optical response, with the quantum metric and Berry curvature governing the symmetric and antisymmetric conductivity components, respectively~\cite{Ahn2022,Moore2025}. The symmetry-based selection of these quantum-geometric contributions and their dependence on the magnetic order parameter remain unexplored. 

In this Letter, we establish a quantum-geometric understanding of the MO response in altermagnetic MnTe by combining first-principles calculations with a Landau symmetry framework. We show that the quantum metric and Berry curvature distinguish MO responses governed by distinct N\'{e}el-order invariants. The quantum metric contributes to the symmetric optical conductivity through the quadratic invariants $(N_x^2-N_y^2,2N_xN_y)$, giving rise to the MOVE with twofold angular dependence. In contrast, the Berry curvature contributes to the antisymmetric Hall conductivity through the cubic invariant $N_y(3N_x^2-N_y^2)$, yielding a threefold angular dependence characteristic of the MOKE. Importantly, the Hall response remains essentially unchanged even when the net ferromagnetic moment is constrained to zero, demonstrating its N\'{e}el-order origin independent of net magnetization. Our findings are corroborated by Landau symmetry analysis, establishing N\'{e}el-order-driven optical responses as key quantum-geometric signatures of unconventional magnetism.

{\it{Quantum geometry and linear optical conductivity}}--- This section introduces the connection between the linear optical conductivity and the quantum geometry, encoded in the quantum geometric tensor. For our purposes, we restrict our discussion to insulating systems, where the optical response is governed entirely by interband transitions. Therefore, the frequency-dependent optical conductivity can then be expressed as \cite{OppeneerBook,Ahn2022,Li2025}, 
\begin{equation}
\sigma_{ab}(\omega)
=-ie^2
\sum_{m,n}\int [d\bm{k}]\,\mathcal{F}_{nm}(\bm{k},\omega)
\Big(g^{ab}_{nm\bm{k}}-\frac{i}{2}\Omega^{ab}_{nm\bm{k}}
\Big),\label{eq:optical_conductivity}
\end{equation}
where $\mathcal{F}_{nm}(\bm{k},\omega)=\frac{f_{nm\bm{k}}\omega_{nm\bm{k}}}
{\hbar\omega_{nm\bm{k}}-\hbar\omega-i\eta}$ and $[d\bm{k}] \equiv \frac{d^3k}{(2\pi)^3}$. The quantities $g^{ab}_{nm\bm{k}}=\frac{1}{2}(r^a_{nm\bm{k}}r^b_{mn\bm{k}}+r^b_{nm\bm{k}}r^a_{mn\bm{k}})$ and $\Omega^{ab}_{mn\bm{k}}=i(r^a_{nm\bm{k}}r^b_{mn\bm{k}}-r^b_{nm\bm{k}}r^a_{mn\bm{k}})$ denote the band-resolved interband quantum metric and Berry curvature matrix elements, respectively. The position operator matrix elements are denoted by $r^a_{nm\bm{k}}=i\langle u_{n\bm{k}}|\partial_{k_a}| u_{m\bm{k}} \rangle$, $f_{nm\bm{k}}=f_{n\bm{k}}-f_{m\bm{k}}$ with $f_{n\bm{k}}$ being the Fermi distribution function, $\hbar \omega_{nm\bm{k}}$ the energy difference between the $n$th and $m$th band, and $\eta$ is the lifetime parameter in  units of energy.

From a symmetry perspective, the quantum metric and Berry curvature possess fundamentally different time-reversal ($\mathcal{T}$) properties. While the quantum metric is $\mathcal{T}$-even, the Berry curvature is $\mathcal{T}$-odd, making broken $\mathcal{T}$ symmetry a necessary condition for a finite Berry curvature but not for a finite quantum metric. Consequently, the quantum metric contributes exclusively to the symmetric optical conductivity, $\sigma^g_{ab}=\sigma^g_{ba}$, whereas the Berry curvature contributes only to the antisymmetric conductivity, $\sigma^\Omega_{ab}=-\sigma^\Omega_{ba}$. Spatial symmetries impose further constraints on these responses. For instance, an $n$-fold rotational symmetry $\mathcal{C}_{nz}$ ($n\geq 3$) enforces $\sigma^g_{xx}=\sigma^g_{yy}$ and $\sigma^g_{xy}=\sigma^g_{yx}=0$. Lowering the rotational symmetry lifts these constraints, allowing $\sigma^g_{xx}\neq\sigma^g_{yy}$ together with finite symmetric off-diagonal components $\sigma^g_{xy}=\sigma^g_{yx}$ \cite{Moore2025}. The anisotropic symmetric optical response therefore provides a direct manifestation of the underlying quantum metric. In contrast, the Berry-curvature response is not constrained by rotational symmetry in this manner but may instead be prohibited by appropriate mirror symmetries \cite{Moore2025}. These contrasting spatial symmetry constraints underpin the distinct optical responses due to the quantum metric and the Berry curvature. In magneto-optics, these two quantum-geometric sectors therefore enter through distinct polarization responses: the anisotropic symmetric conductivity governs the MOVE through the quantum metric, whereas the antisymmetric Hall conductivity gives rise to the MOKE through the Berry curvature. We now make this correspondence explicit by deriving the general MO response in terms of the dielectric tensor and the optical conductivity.

 {\it{Magneto-optical effect}}---
 To characterize the MO response in terms of the dielectric tensor, we derive the general expression for the complex polarization rotation in the polar reflection geometry. For normally incident light propagating along the $z$-direction and linearly polarized in the $xy$ plane at an angle $\alpha$ with respect to the $x$-axis, $\bm{E}_i=E_0 \, e^{i(kz-wt)}\bm{\hat{e}}_i$, with $\bm{\hat{e}}_i=(\cos\alpha, \sin\alpha$), the complex polarization angle is (for detail derivation, see Supplemental Material (SM) \cite{RoojSM2026})
\begin{equation}
    \phi+i\varepsilon={A}/{B},\label{eq:MO}
\end{equation}
where $A=(\epsilon_{xy}-\epsilon_{yx})+(\epsilon_{xx}-\epsilon_{yy})\sin2\alpha-(\epsilon_{xy}+\epsilon_{yx})\cos2\alpha$, $B=(n_++n_-)(1-n_+n_-)-(\epsilon_{xx}-\epsilon_{yy})\cos2\alpha-(\epsilon_{xy}+\epsilon_{yx})\sin2\alpha$, and $\phi$ and $\varepsilon$ are the polarization rotation angle and the ellipticity, respectively. Here, $\bm{\epsilon}$ is the dielectric tensor and $n_\pm$ are the two solutions of the Fresnel equation of the form
\begin{equation}
    n_{\pm} =\sqrt{\frac{\epsilon_{xx}+\epsilon_{yy}}{2}\pm \sqrt{\frac{(\epsilon_{xx}-\epsilon_{yy})^2}{4}+\epsilon_{xy}\epsilon_{yx}}}.\label{eq:MO1}
\end{equation}
Expressing the dielectric tensor components in terms of the optical conductivity tensor components through $\epsilon_{ij}=\delta_{ij}+\frac{4\pi i}{\omega}\sigma_{ij}$ allows the complex polarization rotation to be directly related to the optical conductivity tensor. We decompose the conductivity into its symmetric and antisymmetric parts, $\sigma_{ij}=\sigma^S_{ij}+\sigma^A_{ij}$, with $\sigma^S_{ij}=(\sigma_{ij}+\sigma_{ji})/2$ and $\sigma^A_{ij}=(\sigma_{ij}-\sigma_{ji})/2$. As established above, these two sectors are respectively associated with the quantum metric and Berry curvature, $\sigma^S_{ij}\equiv\sigma^g_{ij}$ and $\sigma^A_{ij}\equiv\sigma^\Omega_{ij}$. In particular, the combinations entering in \cref{eq:MO} separate naturally into
\begin{equation}
\sigma_{xy}+\sigma_{yx}\equiv\sigma^g_{xy}+\sigma^g_{yx},\quad
\sigma_{xy}-\sigma_{yx}\equiv \sigma^\Omega_{xy}-\sigma^\Omega_{yx}\label{eq:MO2},
\end{equation}
while $\sigma_{xx}=\sigma^g_{xx}$ and $\sigma_{yy}=\sigma^g_{yy}$. 
The anisotropic symmetric conductivity, through the combinations $\sigma^g_{xx}-\sigma^g_{yy}$ and $\sigma^g_{xy}+\sigma^g_{yx}$, governs the MOVE, while $\sigma^\Omega_{xy}-\sigma^\Omega_{yx}$ gives rise to the MOKE. Both will appear in reflection; a recent investigation \cite{Li2025} only considered the MOKE response.
The relative contribution of these quantum-geometric quantities to the MO response is dictated by the symmetry of the underlying magnetic state.

\begin{figure}
\includegraphics[width=1\linewidth]{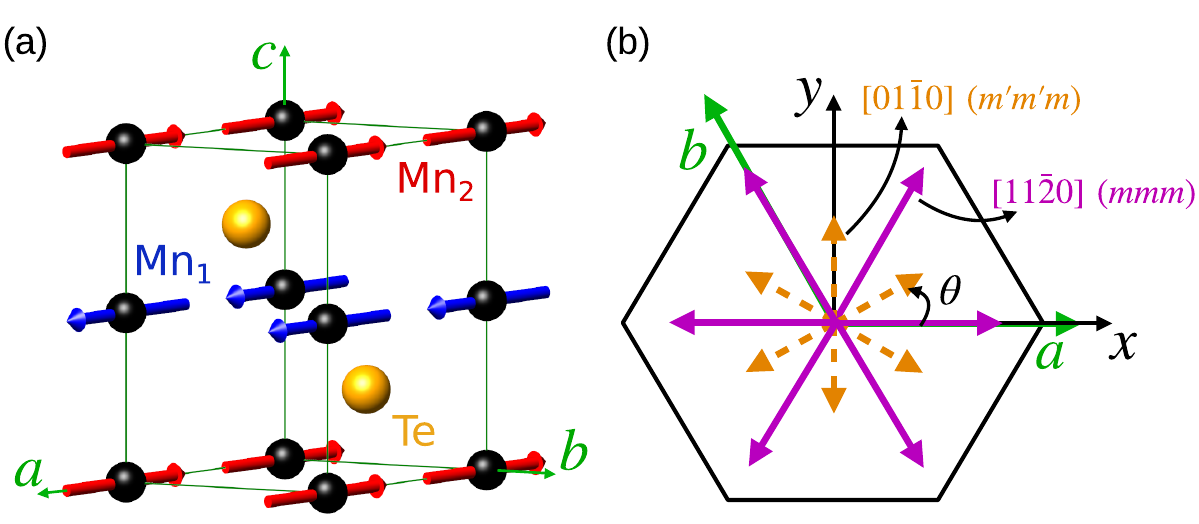}
	\caption{\label{Fig:structure}(a) Crystal and magnetic structures of hexagonal $\alpha$-MnTe. (b) Projection onto the basal ($c=0$) plane illustrating two symmetry-inequivalent N\'{e}el-vector orientations along the magnetic easy axes $[11\bar{2}0]$ and $[01\bar{1}0]$, with the N\'{e}el-vector rotation parameterized by the angle $\theta$ measured relative to the crystallographic $a$ axis.} 
\end{figure}

{\it{Bulk MnTe and symmetry}}--- We now examine the altermagnetic hexagonal bulk $\alpha$-MnTe (point group $D_{6h}$) \cite{RoojMnTe2024, Bilbao2006}, whose crystal and magnetic structures are shown in \cref{Fig:structure}(a). The Mn moments are ferromagnetically aligned within the basal $ab$ planes and antiferromagnetically coupled between adjacent planes along the $c$ axis \cite{Kriegner2016, RoojMnTe2024}. The magnetic configuration is conveniently described by the N\'{e}el vector $\bm{N}=\bm{M}_1-\bm{M}_2$, where $\bm M_1$ and $\bm M_2$ denote the magnetic moments on the two Mn sublattices. Owing to the easy-plane magnetocrystalline anisotropy, $\bm{N}$ is confined to the basal plane \cite{MAE2017}. The combination of hexagonal symmetry and magnetocrystalline anisotropy gives rise to two symmetry-inequivalent orientations of $\bm{N}$, represented by the $[11\bar{2}0]$ ($\theta=60^\circ$) and $[01\bar{1}0]$ ($\theta=90^\circ$) directions, respectively. These directions correspond to the relativistic magnetic point groups (MPGs) $mmm$ and $m'm'm$, with the symmetry-equivalent orientations shown in \cref{Fig:structure}(b) \cite{MnTeAHE2023}. The two MPGs impose distinct constraints on the net magnetization. In particular, the $mmm$ symmetry forbids a net magnetization, while $m'm'm$ permits a finite net magnetization $M_z$ along the $[0001]$ direction \cite{Aroyo:xo5013,MagneticSpaceGroup}. These distinct magnetic symmetries consequently lead to different constraints on the optical conductivity. For both configurations, the absence of an $n$-fold rotational symmetry $\mathcal{C}_{nz}$ ($n\geq 3$) permits an in-plane anisotropic symmetric conductivity, with $\sigma^g_{xx}\neq\sigma^g_{yy}$ and $\sigma^g_{xy}=\sigma^g_{yx}\neq 0$. In contrast, despite broken $\mathcal{T}$ symmetry, the three mutually orthogonal mirror planes of MPG $mmm$ forbid the Berry-curvature-induced antisymmetric Hall conductivity $\sigma^\Omega_{xy}$, while $m'm'm$ permits a finite $\sigma^\Omega_{xy}$. The calculated MO conductivities for a chosen value of $\eta=0.051$~eV are presented in the SM \cite{RoojSM2026}. 

\begin{figure}
\includegraphics[scale=0.383]{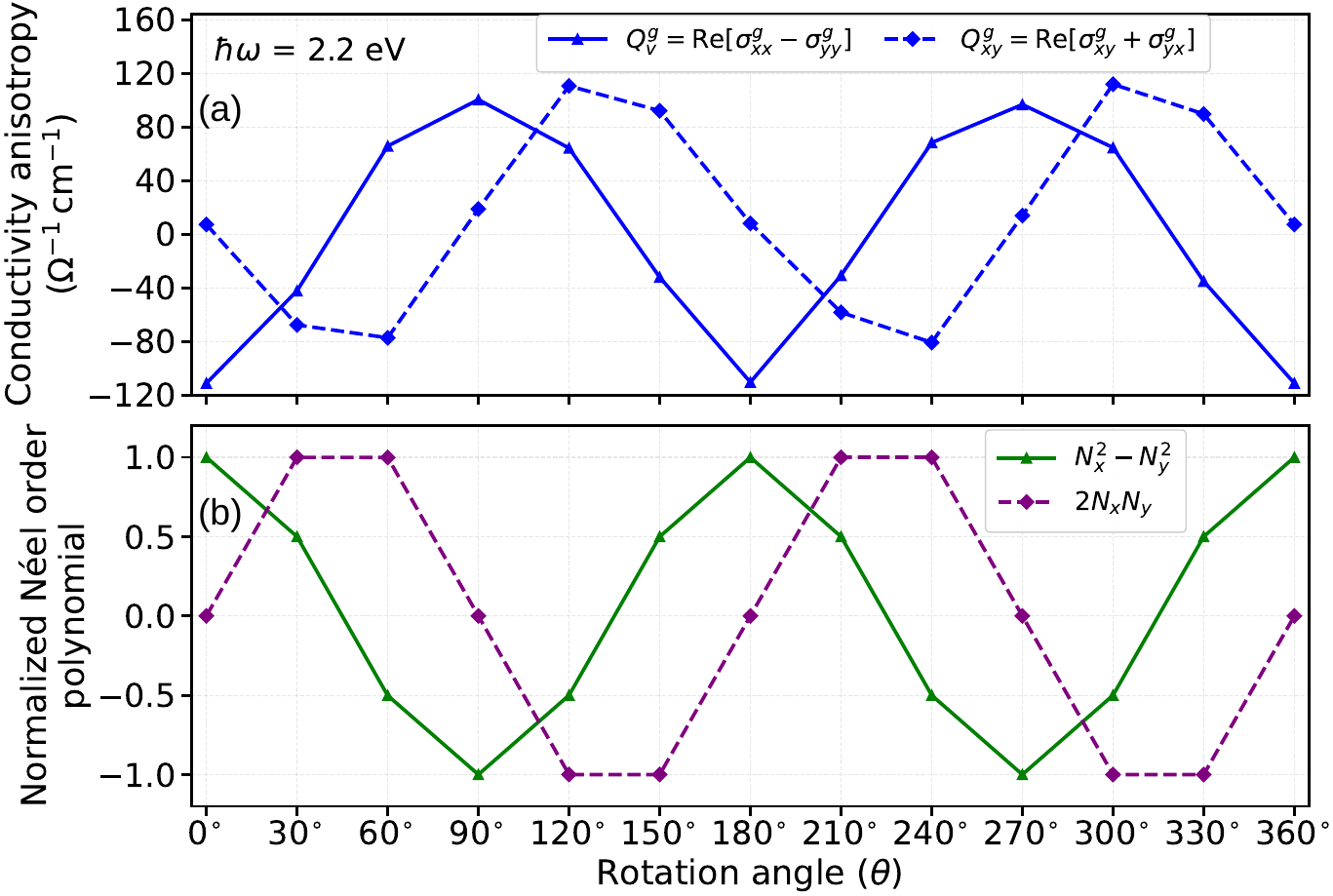}
	\caption{\label{Fig:quadrupole_real}(a) Angular dependence of the symmetric conductivity anisotropy $Q_v^g$ and $Q_{xy}^g$ at a photon energy of 2.2 eV. The in-plane N\'{e}el vector is rotated by an angle $\theta$ from the crystallographic $a$ axis. (b) Quadratic N\'{e}el-order combinations $N_x^2-N_y^2$ and $2N_xN_y$ as functions of $\theta$. $Q_v^g$ and $Q_{xy}^g$ exhibit the same angular dependences as $N_x^2-N_y^2$ and $2N_xN_y$, respectively.}
\end{figure}

{\it{N\'{e}el-order dependence of the MO response}}--- Having established that the relativistic MPGs, and hence
the symmetry-allowed optical conductivity, depends on the orientation of the N\'{e}el vector, we now investigate this dependence explicitly. We stepwise rotate the in-plane N\'{e}el vector,
$(N_x,N_y)=N(\cos\theta,\sin\theta)$, with $\theta$ measured from the crystallographic $a$ axis, as illustrated in \cref{Fig:structure}(b). We first consider the anisotropic symmetric part of the MO response, which is exclusively associated with the quantum-metric. To quantify its anisotropy, we introduce the quantities
\begin{equation}
Q_v^g=\operatorname{Re}[\sigma^g_{xx}-\sigma^g_{yy}],\qquad
Q_{xy}^g=\operatorname{Re}[\sigma^g_{xy}+\sigma^g_{yx}].\label{eq:quadrupole}
\end{equation}
These two quantities constitute the anisotropic symmetric conductivity entering the MOVE and therefore provide the relevant optical observables for resolving the symmetry of the quantum-metric contribution. Figure~\ref{Fig:quadrupole_real}(a) shows that the calculated $Q_v^g$ and $Q_{xy}^g$ exhibit $\cos2\theta$ and $\sin2\theta$ angular dependences, respectively. These quantities follow the same angular symmetry as the quadratic combinations $N_x^2-N_y^2$ and $2N_xN_y$, constructed from the self-consistently obtained N\'{e}el vector, as shown in  \cref{Fig:quadrupole_real}(b). This correspondence demonstrates that the quantum-metric induced anisotropic conductivity carries a direct symmetry fingerprint of the N\'{e}el order. The imaginary part of the MO conductivity also exhibits the same angular dependence with the N\'{e}el-vector rotation, shown in the SM \cite{RoojSM2026}. 

\begin{figure}
\includegraphics[scale=0.35]{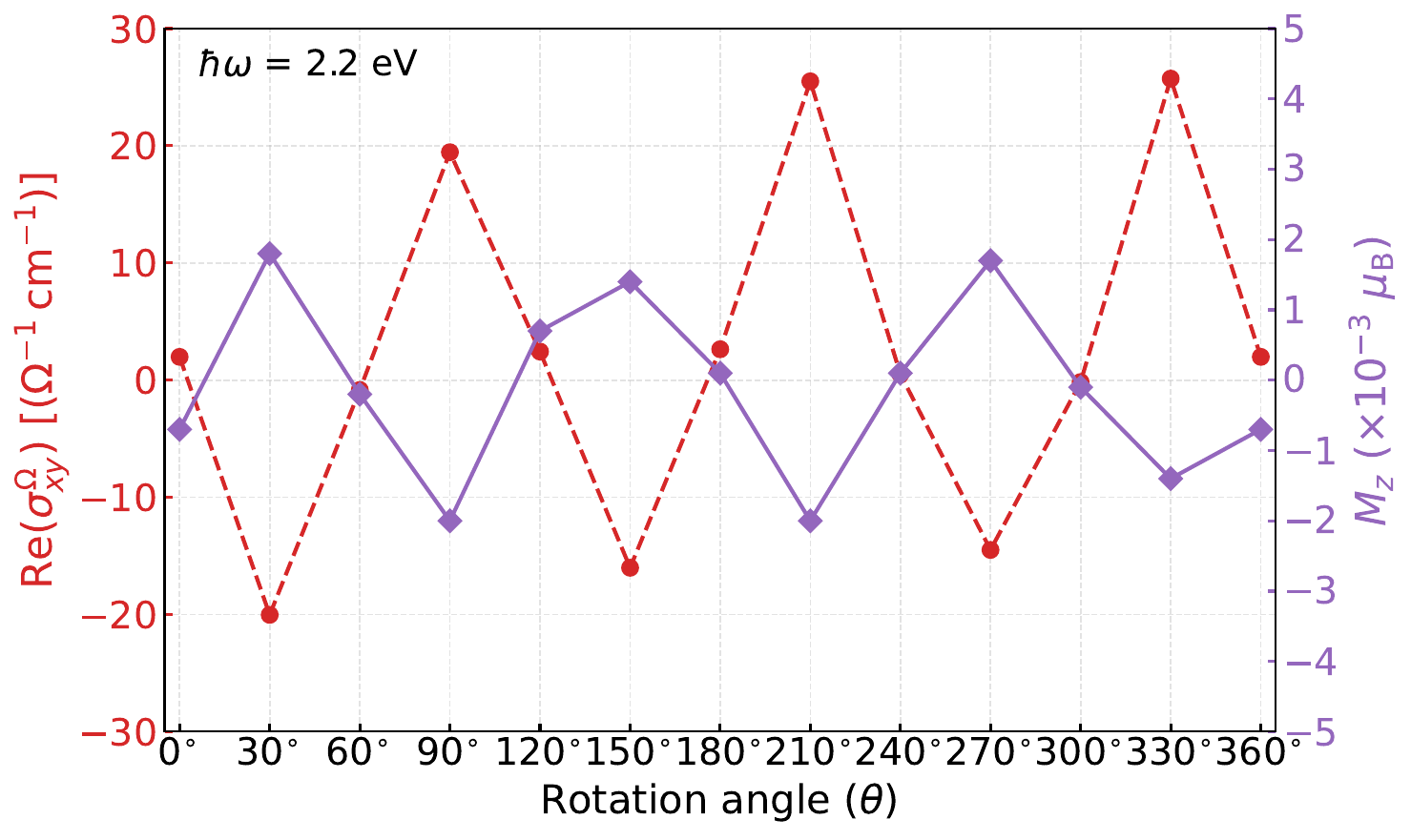}
	\caption{\label{Fig:MzSigma}Dependence of the real part of the Berry-curvature-induced Hall conductivity $\sigma^\Omega_{xy}$ (red) and the out-of-plane magnetization $M_z$ (purple) on the in-plane N\'{e}el vector angle $\theta$. }
\end{figure}

\begin{figure}
\includegraphics[scale=0.28]{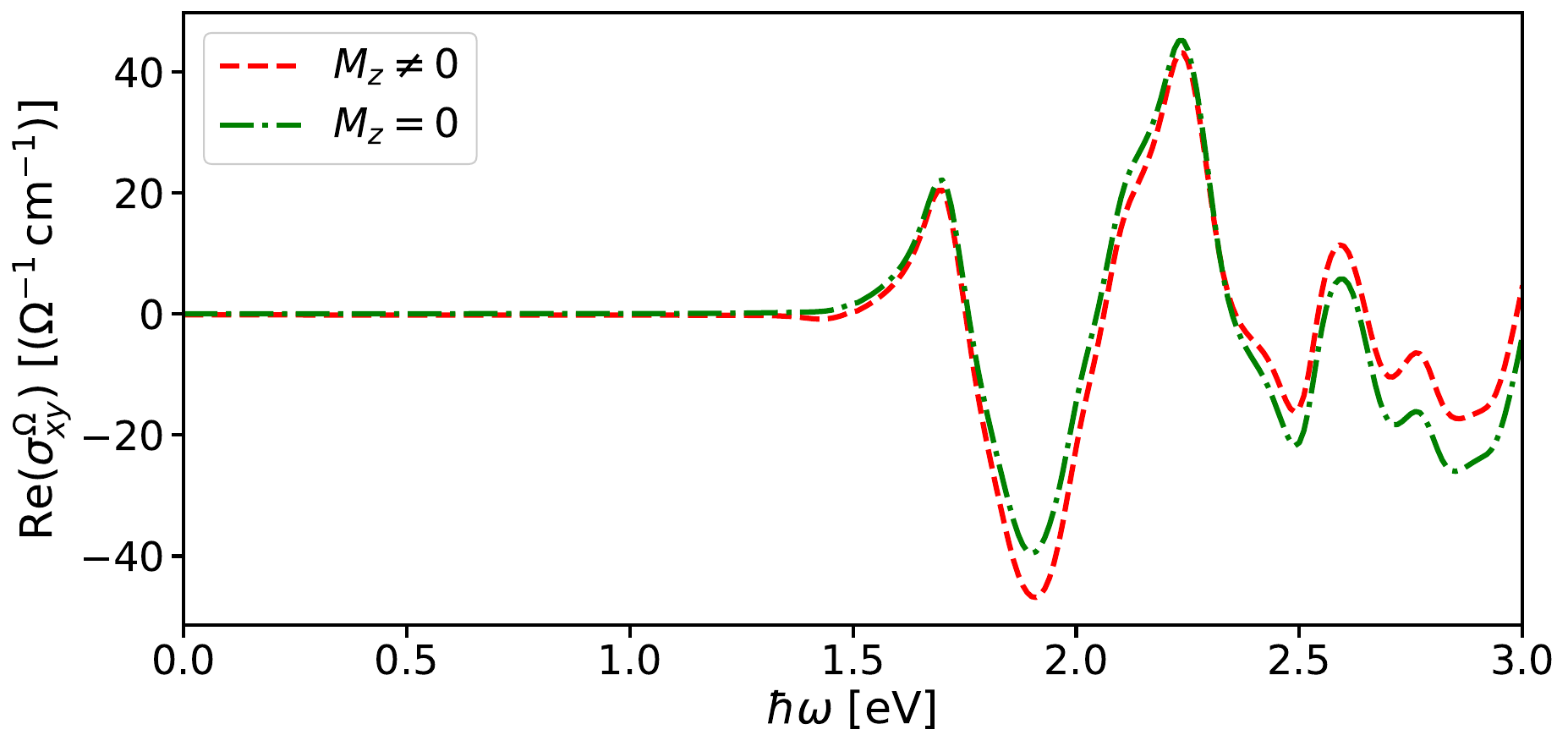}
	\caption{\label{Fig:sigmaComparison}Real part of the Berry-curvature-induced Hall conductivity $\sigma^\Omega_{xy}$ for $\bm N\parallel[01\bar{1}0]$ with (red) and without (green) the out-of-plane canting. The calculations with the self-consistent $M_z$ and with $M_z=0$ show nearly identical frequency dependence.}
\end{figure}

The antisymmetric conductivity, i.e., the Berry-curvature-induced Hall conductivity $\sigma^\Omega_{xy}$, exhibits a qualitatively distinct angular dependence as the N\'{e}el vector is rotated. As discussed above, symmetry suggests that $\sigma^\Omega_{xy}$ vanishes for the $mmm$ configuration but remains finite for the $m'm'm$ configuration [see, e.g., \cref{Fig:structure}(b)].  Figure~\ref{Fig:MzSigma} shows that $\sigma^\Omega_{xy}$ follows a $\sin 3\theta$ dependence, analogous to that of the out-of-plane net magnetization $M_z$, with the small residual values at the symmetry-forbidden $mmm$ orientations lying within the numerical accuracy of the calculation. This correlation is consistent with the conventional relation $\sigma^\Omega_{xy}\propto M_z$ and, at this stage, suggests that the weak ferromagnetic moment may provide the relevant Hall response. To determine whether the Hall response requires a finite weak-ferromagnetic moment, we consider the limiting case in which the magnetic moments are constrained to their initial directions, thereby suppressing spin canting and imposing $M_z=0$. Remarkably, suppressing the out-of-plane magnetization produces only negligible changes in $\sigma^\Omega_{xy}$ over the entire frequency range, as shown in \cref{Fig:sigmaComparison}. Thus, although $M_z$ and $\sigma^\Omega_{xy}$ exhibit the same angular dependence, the weak ferromagnetic moment is not responsible for the Hall response in MnTe. Instead, the persistence of the Hall conductivity in the absence of $M_z$ points to a direct connection to the underlying N\'{e}el order.

These results have a direct consequence for the MO response. From \cref{eq:MO}, the complex polarization rotation contains the anisotropic symmetric combinations
$\sigma^g_{xx}-\sigma^g_{yy}$ and
$\sigma^g_{xy}+\sigma^g_{yx}$, which govern the MOVE, as well as the antisymmetric combination
$\sigma_{xy}^\Omega-\sigma_{yx}^\Omega$, which gives rise to the MOKE. The calculated angular dependence therefore demonstrate that the two quantum-geometric contributions to the MO response carry distinct signatures of the same underlying N\'{e}el order:
a twofold dependence in the symmetric sector and a threefold dependence in the antisymmetric sector.

The appearance of these well-defined angular forms raises a natural question: can both MO responses be expressed directly in terms of symmetry-allowed polynomials of the N\'{e}el-order parameter, without invoking the microscopic details of the electronic structure? To answer this question, we formulate the MO response within Landau phenomenology and identify the lowest-order N\'{e}el-order combinations transforming according to the symmetry of each conductivity channel.

{\it{Landau theory}}--- The distinct angular dependence of the symmetric and antisymmetric MO responses suggest that they originate from different symmetry-allowed combinations of the magnetic order parameters of the underlying magnetic structure. We therefore use
Landau phenomenology to identify the lowest-order polynomial combinations of the order parameters that transform according to the symmetry of the corresponding optical responses. Rather than constructing the Landau free energy explicitly, we use its underlying symmetry principle to establish the symmetry allowed couplings between the magnetic order and the optical conductivity. This provides a direct connection between the angular dependence of the MO conductivity and the symmetry of the magnetic order parameters.

For hexagonal MnTe, the relevant magnetic order parameters are the net magnetization $\bm M=\bm M_1+\bm M_2$ and the N\'{e}el vector $\bm N=\bm M_1-\bm M_2$, where $\bm M_1$ and $\bm M_2$ are the magnetic moments of two sublattices [e.g., see \cref{Fig:structure}(a)]. The magnetization transforms as an axial vector irreducible representation (IR) under the parent $D_{6h}$ point group and therefore decomposes as $\Gamma_A=A_{2g}\oplus E_{1g}$, where $M_z\in A_{2g}$ and $(M_x,M_y)\in E_{1g}$. As established in the previous work, the altermagnetic spin splitting in hexagonal MnTe is described by the polynomial of the form $k_y k_z(3k_x^2-k_y^2)$  \cite{RoojArxiv2025}, which transforms according to the $B_{1g}$ IR \cite{Kreisel}. Consequently, the N\'{e}el vector $\bm{N}$ transforms as the direct product of the axial vector IR and the altermagnetic spin splitting IR, i.e., \cite{LandauPaul,Kreisel}
\begin{equation}
    \Gamma_N=(A_{2g}\oplus E_{1g})\otimes B_{1g}=B_{2g}\oplus E_{2g}\label{eq:Landau2} \,.
\end{equation}
Experimentally, the N\'{e}el vector in hexagonal MnTe is confined to the basal plane and is therefore described by $\bm{N}=N_x\hat{x}+N_y\hat{y}$. Accordingly, only the in-plane components $(N_x,N_y)$ are relevant, and they transform according to the two-dimensional IR $E_{2g}$.

We first consider the symmetric part of the linear MO
conductivity. For an in-plane N\'{e}el vector, the relevant conductivity tensor components are $\sigma^g_{xx}$, $\sigma^g_{yy}$, $\sigma^g_{xy}$, and $\sigma^g_{yx}$, which originate from the quantum metric. Since the symmetric conductivity is even
under $\mathcal{T}$ symmetry, its magnetic-order dependence must be described by the even polynomials in the order parameters \cite{Kimel2024}.
For the in-plane N\'{e}el vector, the lowest-order combinations are quadratic and transform according to the symmetric tensor power of the IR $E_{2g}$ (see SM \cite{RoojSM2026})
\begin{equation}
    \operatorname{Sym}^{2}(E_{2g})=A_{1g}\oplus E_{2g} \,.\label{eq:Landau3}
\end{equation}
The totally symmetric $A_{1g}$ representation corresponds to the
isotropic invariant $ N_x^2+N_y^2$ while the $E_{2g}$ representation is spanned by the anisotropic basis
functions
\begin{equation}
    \left(N_x^2-N_y^2,\,2N_xN_y\right).
    \label{eq:Landau4}
\end{equation}
An analogous decomposition can be constructed from the in-plane
magnetization, $(M_x,M_y)\in E_{1g}$. However, the equilibrium state of hexagonal MnTe has vanishing in-plane magnetization for all in-plane N\'{e}el vector orientations considered here. Consequently, the in-plane magnetization does not contribute to the present symmetry analysis. 

The anisotropic components of the symmetric conductivity, $Q_v^g$ and $Q_{xy}^g$ [Eq.~\eqref{eq:quadrupole}], form an $E_{2g}$ doublet (see SM \cite{RoojSM2026}). Matching them with the quadratic $E_{2g}$ basis in \cref{eq:Landau4} provides a direct connection such that
\begin{equation}
    Q_v^g\propto N_x^2-N_y^2, \quad Q_{xy}^g\propto 2N_xN_y.
    \label{eq:Landau6}
\end{equation}
For fixed $|\bm N|$, these relations in \cref{eq:Landau6} correspond to the characteristic $\cos2\theta$ and $\sin2\theta$ dependence observed in our numerical calculations as evident from \cref{Fig:quadrupole_real}. The same symmetry is obtained for the imaginary part of the symmetric conductivity, as shown in the SM \cite{RoojSM2026}. Thus, the anisotropic symmetric conductivity that governs MOVE transforms according to the same $E_{2g}$ representation as the quadratic N\'{e}el order combinations, providing a direct symmetry fingerprint of the antiferromagnetic state.

We next consider the antisymmetric part of the MO conductivity. Unlike the symmetric conductivity, the Berry-curvature-induced Hall response is odd under $\mathcal{T}$-symmetry and therefore can couple only to odd-order combinations of the magnetic order parameters. The out-of-plane magnetization transforms as
$M_z\in A_{2g}$, and the antisymmetric Hall conductivity
$\sigma^\Omega_{xy}=-\sigma^\Omega_{yx}$ belongs to the same IR $A_{2g}$.
A linear coupling to the in-plane N\'{e}el vector is forbidden because $E_{2g}$ does not contain the $A_{2g}$ representation. The leading N\'{e}el-order contribution to the Hall response must therefore be cubic. By imposing the generators of the $D_{6h}$ point group, we obtain the cubic polynomials constructed from the in-plane N\'{e}el vector of the form $N_y(3N_x^2-N_y^2)$ (see SM \cite{RoojSM2026} for derivation).
Since this cubic N\'{e}el-order invariant, $M_z$, and $\sigma^\Omega_{xy}$, all transform according to $A_{2g}$ IR, symmetry
permits the couplings
\begin{equation}
    \sigma^\Omega_{xy}
    \propto N_y(3N_x^2-N_y^2),\quad M_z\propto N_y(3N_x^2-N_y^2).
    \label{eq:Landau8}
\end{equation}

The calculated angular dependence of $M_z$ and
$\sigma^\Omega_{xy}$ shown in \cref{Fig:MzSigma} follow this characteristic threefold form, in agreement with the cubic N\'{e}el-order invariant. Although $M_z$ and $\sigma^\Omega_{xy}$ exhibit the same angular dependence, this correspondence does not imply that the weak-ferromagnetic moment is the microscopic origin of the Hall response. Indeed, as demonstrated in \cref{Fig:sigmaComparison}, the Berry-curvature Hall conductivity remains essentially unchanged when the out-of-plane canting is suppressed and $M_z$ is constrained to zero. This demonstrates that the Hall response, which gives rise to the MOKE, admits a direct symmetry-allowed coupling to the cubic N\'{e}el-order invariant in \cref{eq:Landau8}, independent of a finite $M_z$.

Taken together, the symmetry analysis establishes a direct correspondence between the N\'{e}el order and the two quantum-geometric sectors of the MO response. The symmetric quantum-metric contribution couples to the quadratic $E_{2g}$ structure $(N_x^2-N_y^2,2N_xN_y)$, producing the characteristic twofold angular dependence, while the antisymmetric Berry-curvature contribution couples to the cubic $A_{2g}$ invariant $N_y(3N_x^2-N_y^2)$, yielding the qualitative characteristic threefold dependence. These results demonstrate that the orientation of the N\'{e}el order is directly encoded in distinct symmetry channels of the MO conductivity.

{\it{Conclusions}}--- Our results establish that the symmetry of N\'{e}el order in altermagnetic MnTe is directly imprinted on its magneto-optical response through quantum geometry. This finding is distinct from that of the recent work of Yang \textit{et al.}\ \cite{Ultrafast2026}, who reported MO Kerr and Voigt responses of MnTe, but did not include the symmetric off-diagonal component  $\sigma^g_{xy}+\sigma^g_{yx}$ that constitutes the quantum metric. We find that the quantum metric produces a twofold optical anisotropy that governs the MOVE, while the Berry curvature generates a threefold Hall response underlying the MOKE. These two responses therefore provide complementary optical hallmarks of the same antiferromagnetic order, despite the absence of a net in-plane magnetization. Most importantly, our calculations demonstrate that the Berry-curvature Hall response in MnTe does not originate from the weak-ferromagnetic moment: it remains essentially unchanged when the out-of-plane moment is constrained to zero. Our results thus reveal quantum geometry as a direct bridge between the magnetic symmetry and optical response of MnTe, and demonstrate that magneto-optics can provide a sensitive probe of its N\'{e}el order. Although our analysis focuses on MnTe, the underlying symmetry-based approach is not limited to this compound and may also be applicable to other centrosymmetric altermagnets. 

{\it{Acknowledgments}}---
 This work has been supported by the Swedish Research Council
(VR) and the Knut and Alice Wallenberg Foundation (Grant Nos.\
2022.0079 and 2023.0336). The calculations were supported by resources provided by the National
Academic Infrastructure for Supercomputing in Sweden (NAISS) at
NSC Link\"oping, partially funded by VR through Grant Nos.\
2022-06725 and 2026-05211.

{\it{Data availability}}--- The data that support the findings of this article are available from the authors upon reasonable request.

%

\end{document}